\documentclass[printer]{aa}  

\usepackage{graphicx}
\usepackage{txfonts}
\begin{document} 
   \title{The dynamics and observability of circularly polarized kink waves}
   \author{N. Magyar
          \inst{1} \fnmsep\thanks{FWO-Vlaanderen fellow.},
          T. Duckenfield
          \inst{1},
          T. Van Doorsselaere
          \inst{1},
          V. M. Nakariakov
          \inst{2,3}          
          }
   \institute{             Centre for mathematical Plasma Astrophysics (CmPA),
KU Leuven, Celestijnenlaan 200B bus 2400, B-3001 Leuven, Belgium \\
              \email{norbert.magyar@kuleuven.be}
         \and
Centre for Fusion, Space and Astrophysics, Physics Department, University of Warwick, Coventry CV4 7AL, UK
         \and 
             School of Space Research, Kyung Hee University, Yongin, 17104, Republic of Korea
           }
 
\abstract
   {Kink waves are routinely observed in coronal loops. Resonant absorption is a well-accepted mechanism that extracts energy from kink waves. Nonlinear kink waves are know to be affected by the Kelvin-Helmholtz instability. However, all previous numerical studies consider linearly polarized kink waves.}
   {We study the properties of circularly polarized kink waves on straight plasma cylinders, for both standing and propagating waves, and compare them to the properties of linearly polarized kink waves.}
   {We use the code \texttt{MPI-AMRVAC} to solve the full 3D Magnetohydrodynamic (MHD) equations for a straight magnetic cylinder, excited by both standing and propagating circularly polarized kink ($m=1$) modes.}
   {The damping due to resonant absorption is independent of the polarization state. The morphology or appearance of the induced resonant flow is different for the two polarizations, however, there are essentially no differences in the forward-modeled Doppler signals. For nonlinear oscillations, the growth rate of small scales is determined by the total energy of the oscillation rather than the perturbation amplitude. We discuss possible implications and seismological relevance.}
   {}

\titlerunning{Circularly polarized kink}
\authorrunning{N. Magyar et al.}

\maketitle

\section{Introduction}

The outer atmosphere of the Sun, the solar corona is a highly structured and dynamic, hot and rarefied plasma. The observed structures are generally thought to outline magnetic field lines, appearing as closed magnetic elements known as coronal loops in active regions, and as plumes in expanding open field lines. Still, the observed fine field-aligned density filamentation of the coronal plasma \citep[e.g.,][]{2014ApJ...788..152R,2020ApJ...892..134W} remains a puzzle, being linked to the detailed mechanism of the coronal heating problem.  
Magnetohydrodynamic (MHD) waves are omnipresent and routinely observed in these structures, both standing and propagating \citep[e.g.][]{2020ARA&A..58..441N}. These waves are of great scientific interest as they might play a role in coronal heating \citep[e.g.,][]{2015RSPTA.37340261A,2020SSRv..216..140V}. On the other hand, properties of the observed waves, such as wave speeds, periods, and so on, allow us to infer properties of the medium in which they propagate, a field known as coronal seismology \citep[e.g.,][]{2012RSPTA.370.3193D,2018AdSpR..61..655A}. Oscillating displacements of coronal loops, interpreted as kink waves, were the first modes used for coronal seismology and are intensively studied since. 
Kink modes are observed as impulsively triggered large amplitude standing waves or small amplitude decayless standing or ubiquitously propagating waves. The strong damping of observed large amplitude standing kink waves initially constituted a puzzle, as dissipative coefficients in the corona appear to be orders of magnitude smaller than required. Resonant absorption, the conversion of the global kink wave energy to localized Alfv\'en waves, was proposed as the main mechanism to explain the damping \citep[e.g.][]{2002ApJ...577..475R,2002ESASP.505..137G}. Recently, evidence on the amplitude-dependent damping of standing kink waves has emerged \citep{2016A&A...590L...5G,2019ApJS..241...31N,2021ApJ...915L..25A}. Resonant absorption, a linear theory, cannot account for this amplitude dependence. Nonlinear mechanisms, such as the Kelvin-Helmholtz instability, or the initiation of a nonlinear azimuthal cascade have been proposed to explain this dependence \citep[][]{2021ApJ...910...58V}. Nevertheless, resonant absorption continues to represent a viable mechanism to explain at least part of the observed damping. Analytical and numerical models building upon the original cylindrical model of \citet{1975...37..3} and \citet{1983SoPh...88..179E} have greatly improved our understanding of kink oscillations in coronal loops, including numerous effects related to loop geometry, structuring, and time-dependent backgrounds, among others. \citep[see, e.g.,][for reviews on the subject]{2009SSRv..149..199R,2021SSRv..217...73N}\par 
Another property of kink waves is their polarization. Most of the existing works focus on the linearly polarized kink wave, even though in analytical studies a circular polarization is assumed, as the mathematical treatment is more convenient in this case. One could argue that this is sufficient, as in the linear regime any state of polarization can be constructed from either linear or circularly polarized waves. However, some properties of the circularly polarized kink waves are not obvious, such as their resonant absorption. Additionally, the nonlinear evolution of circularly polarized kink waves might be different from their linearly polarized counterpart. In observations, standing kink waves are mostly categorized as `vertically' or `horizontally' polarized, meaning linear polarization with the displacement in or out of the plane of the oscillating loop, respectively. Nevertheless, using three-dimensional (3D) reconstructions from the STEREO EUVI/A and B spacecraft, \citet{2009SSRv..149...31A} found that the loop oscillation classified as linearly polarized horizontal oscillation by \citet{2008A&A...489.1307W} using two-dimensional (2D) imaging, was in fact a circularly polarized oscillation. Moreover, \citet{2008A&A...489.1307W} found that out of the 14 loop oscillations studied, more than half of the cases were not classifiable as either horizontally or vertically polarized, which might be due to more complicated motions, e.g., elliptical or circular polarization. Therefore, a significant proportion of observed kink oscillating loops could be elliptically or circularly polarized. Circularly polarized, propagating kink waves were observed also in chromospheric magnetic elements \citep{2017ApJ...840...19S}. Theoretically, a circular polarization of kink waves is expected to evolve in loops with twisted magnetic fields, even if the initial perturbation is linearly polarized \citep{2015A&A...580A..57R}. Although not transverse but slow type wave, it is worth mentioning that \citet{2017ApJ...842...59J} observed a circularly polarized m=1 mode showing a spiral pattern in a sunspot. \par 
In this paper, we study the circularly polarized kink oscillations, both standing and propagating, of a straight magnetic cylinder, for both linear and nonlinear perturbation amplitudes. The paper is organized as follows. In Section~\ref{two}, we present the numerical method and describe the numerical models. In Section~\ref{three}, we present the results, along with some discussion. Finally, in Section~\ref{four}, we summarize the results and present some possibilities for future work. 

\section{Numerical method and model} \label{two}

We run ideal 3D MHD simulations with \texttt{MPI-AMRVAC}\footnote{http://amrvac.org/} \citep{2014ApJS..214....4P,2018ApJS..234...30X}. The second-order \texttt{tvd} solver is used with \texttt{woodward} slope limiter, unless otherwise specified. The numerical setup at equilibrium consists of a straight cylindrical flux tube embedded in a uniform atmosphere. The background magnetic field $B_0 = \mathrm{10\ G}$ is straight and uniform throughout the domain, and along the $z$-axis. The flux tube is defined by its three times higher density $\rho_i$, than that of the background, $\rho_0 \approx 2.3 \cdot 10^{-12}\ \mathrm{kg\ m^{-3}}$. The inhomogeneous layer between the inside and outside of the flux tube is defined as a sinusoidal variation in density, with different widths $l$. Then, the density in the whole domain is defined as:
\begin{equation}
    \rho(r) = 
    \begin{cases}
      &\rho_i, r < a - l/2,\\
      &\rho_i - (\rho_i - \rho_0)\mathrm{sin}\left(\frac{\pi}{2 l}(r - (a-l/2)) \right), a + l/2 \leq r \geq a - l/2, \\
      &\rho_0, r > a + l/2\\
    \end{cases}
\end{equation}
Gravity and non-ideal effects, such as optically thin radiative losses and thermal conduction are not included in this model. On top of the background equilibrium we impose velocity and magnetic field perturbations of the kink wave solution \citep[e.g.,][]{1975...37..3,1983SoPh...88..179E,2020ApJ...899..100V}, for studying the circularly polarized standing kink wave. In Cartesian coordinates, the solution reads:
\begin{equation}
  \begin{array}{cc}
    \mathbf{v}_x =& -\frac{1}{\sqrt(2)} A \frac{2 L}{\pi B_0} v_x(r) \mathrm{sin}\left(\frac{\pi}{L} z\right) ,\\
    \mathbf{b}_x =& A \frac{2 L}{\pi B_0} b_x(r)  \mathrm{cos}\left(\frac{\pi}{L} z\right),\\
    \mathbf{v}_y =& \frac{1}{\sqrt(2)} A \frac{2 L}{\pi B_0} v_y(r) \mathrm{sin}\left(\frac{\pi}{L} z\right),\\
    \mathbf{b}_y =& A \frac{2 L}{\pi B_0} b_y(r)  \mathrm{cos}\left(\frac{\pi}{L} z\right),\\
    \mathbf{v}_z =& 0,\\
    \mathbf{b}_z =& \frac{1}{B_0} A b_z(r)\mathrm{sin}\left(\frac{\pi}{L} z\right) ,
  \end{array}
  \label{pert1}
\end{equation}
where
\begin{equation}
  \begin{array}{ll}
  v_x(r) =& 
  \begin{cases} 
     \frac{\frac{k_i x^2 \left(J_0\left(k_i r \right)-J_2\left(k_i
   r \right)\right)}{r^2}+\frac{2 y'^2 J_1\left(k_i
   r \right)}{r^3}}{2 J_1(k_i a)}, r\leq  a,\\
      \frac{\frac{2 y'^2 K_1\left(k_e r \right)}{r^3}-\frac{k_e x^2
   \left(K_0\left(k_e r\right)+K_2\left(k_e r \right)\right)}{r^2}}{2
   K_1(k_e a)}, r> a,\\
 \end{cases} \\
  v_y(r) =& 
   \begin{cases} 
      \frac{x y' \left(k_i r \left(J_0\left(k_i r \right)-J_2\left(k_i
   r \right)\right)-2 J_1\left(k_i r \right)\right)}{2 r^3
   J_1(k_i a)},  r \leq  a,\\
      -\frac{x y' \left(2 K_1\left(k_e r \right)+k_e r \left(K_0\left(k_e
   r \right)+K_2\left(k_e r \right)\right)\right)}{2 r^3
   K_1(k_e a)}, r > a,\\
 \end{cases}  \\
  b_x(r) =& 
  \begin{cases} 
     \frac{x y' \left(2 J_1\left(k_i r \right)- k_i r \left(J_0\left(k_i
   r \right)-J_2\left(k_i r \right)\right)\right)}{2 r^3
   J_1(k_i a)}, r\leq  a,\\
     \frac{x y' \left(2 K_1\left(k_e r \right)+ k_e r \left(K_0\left(k_e
   r \right)+K_2\left(k_e r \right)\right)\right)}{2 r^3
   K_1(k_e a)}, r > a,\\
 \end{cases} \\
  b_y(r) =& 
   \begin{cases} 
      -\frac{\frac{2 x^2 J_1\left(k_i r\right)}{r^3}+\frac{k_i y'^2
   \left(J_0\left(k_i r \right)-J_2\left(k_i r \right)\right)}{r^2}}{2
   J_1(k_i a)},  r \leq a,\\
      \frac{\frac{k_e y'^2 \left(K_0\left(k_e r \right)+K_2\left(k_e
   r \right)\right)}{r^2}-\frac{2 x^2 K_1\left(k_e
   r \right)}{r^3}}{2 K_1(k_e a)},  r > a,\\
 \end{cases}  \\
  b_z(r) =& 
   \begin{cases} 
      \frac{x}{r} \frac{J_1\left(k_i r \right)}{J_1(k_i a)},  r \leq a,\\
      \frac{x}{r} \frac{K_1\left(k_e r \right)}{K_1(k_e a)},  r > a,\\
 \end{cases} 
     \label{pert2}
  \end{array}
\end{equation}
with $r = \sqrt{x^2 + y'^2}$ and $a = 1\ \mathrm{Mm}$ is the radius of the tube. Note that $y' = 0$ denotes the center of the displaced flux tube along $x = 0$ and not the center of the flux tube in equilibrium. The initial displacement of the flux tube is required as a tube undergoing a circularly polarized kink wave does not cross the equilibrium position. $A$ is the amplitude of the perturbation, in units of pressure. Here we use units in which the magnetic permeability $\mu = 1$. The displacement is given by:
\begin{equation}
    y' = y -  A \frac{2 L^2}{a \pi ^2 B_0^2}\mathrm{sin}(\frac{\pi}{L} z). 
\end{equation}
Note that while the displacement is in the $y$-direction, the initial velocity perturbation inside the flux tube is in the negative $x$-direction. The coefficients $k_{i} = k_{e} \approx  0.0112$ are the fundamental radial wave numbers. $J_i$ is the Bessel function of first kind and $K_i$ is the modified Bessel function of second kind. $L = 200\ \mathrm{Mm}$ is the length of the flux tube from foot point to foot point. $A$ is the amplitude of the initial perturbation, in units of pressure, and $v_z$ is neglected due to a sufficiently low plasma beta, $\beta \approx 0.2$ \citep{2016ApJS..223...23Y,2020ApJ...899..100V}. \par
For studying the circularly polarized standing kink wave, the symmetrical properties of the standing mode are exploited in order to halve the computational costs. In this sense, the spatial extents are $[-3.5a,3.5a]^2 \times [0,L/2]$, where $L/2$ denotes the position of the anti-node of the fundamental kink. The same spatial extent is used also for studying propagating waves. The boundary conditions are set `open' or continuous, zero divergence laterally, while the foot points are fixed by forcing the velocity components to be anti-symmetric at the lower $z$-axis boundary. At the top $z$-axis boundary, the fundamental mode is mirrored by setting $v_z, b_x, b_y$ anti-symmetric, and the other variables symmetric. Simulations are run for both the standing and propagating mode. For propagating waves, at the bottom $z$-axis boundary we employ a driver of circularly polarized kink waves, by setting: 
\begin{equation}
         v_x(t) = A\ \mathrm{sin}\left(\frac{2 \pi}{P} t \right), \quad
         v_y(t) = A\ \mathrm{cos}\left(\frac{2 \pi}{P} t \right),
\end{equation}
where $P \approx 86\ \mathrm{s}$ is the wave period. Note that for propagating waves the driven wave period is shorter than the fundamental standing mode period, in order for the kink wave to evolve significantly by the time it reaches the top $z$-axis boundary. For simulating propagating waves the top $z$-axis boundary is set to be continuous or zero divergence. \par 
The numerical domain consists of $256^2 \times 64$ cells of uniform resolution, with fewer cells along the $z$-direction, along which the solution is expected to be smooth. We have conducted a series of convergence studies with both lower and higher resolution, and qualitative differences can be observed for high-amplitude runs which show nonlinear small-scale generation, as expected.
Stemming from the piecewise nature of the kink solution in Eqs.~\ref{pert2} combined with the discretized numerical grid, a non-zero magnetic field divergence is generated at the flux tube boundary as initial condition. Therefore, in order to eliminate the initial divergence and to maintain it at low levels, we employ the \texttt{multigrid} method of \citet{2019CoPhC.24506866T}, which uses a projection scheme to remove the divergence part of the magnetic field.    

\section{Results and Discussion} \label{three}

\begin{figure*}[h]
    \centering     
        \includegraphics[width=0.75\textwidth]{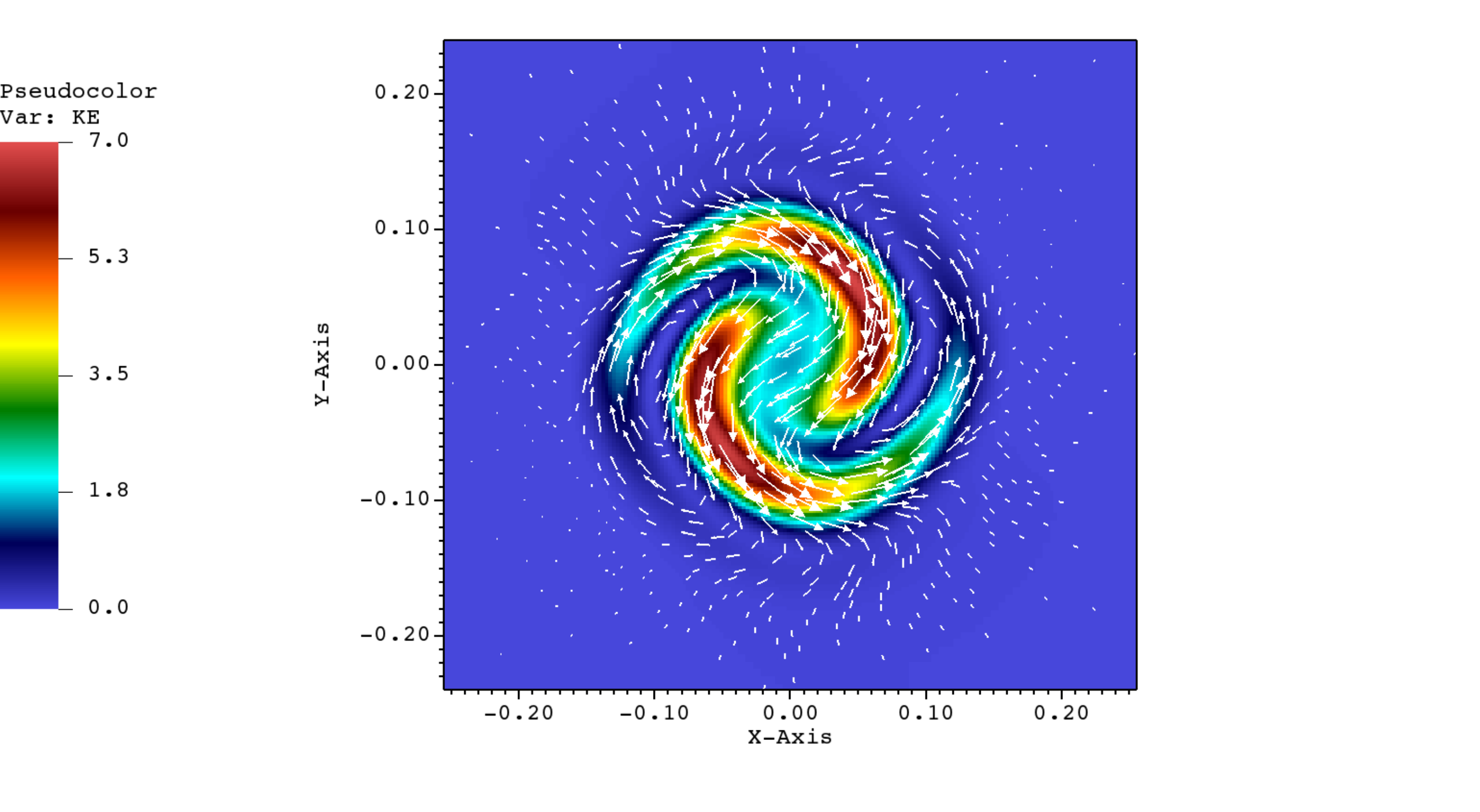} 
        \caption{Plot of kinetic energy and velocity vectors in the cross-sectional slice at the midpoint of the flux tube, for the circularly polarized standing kink mode, after approximately two fundamental kink mode periods of evolution. Axis units are 10 Mm. Kinetic energy is in code units. The size of the velocity vectors is proportional to their magnitude. (An animation of this figure is available in the online version)}
        \label{fig1}
\end{figure*}

\begin{figure*}[h]
    \centering     
        \includegraphics[width=0.75\textwidth]{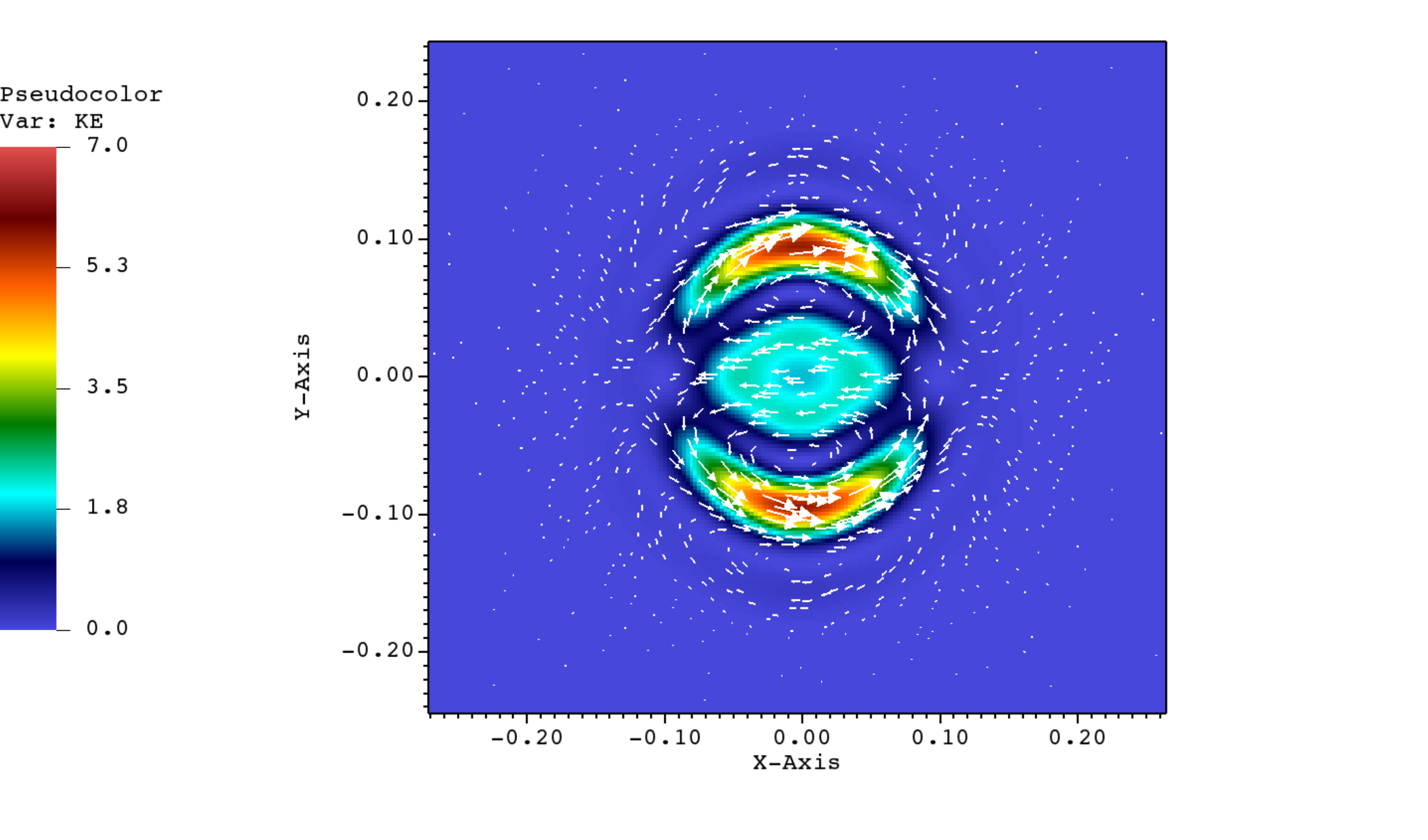} 
        \caption{The same as in Fig.~\ref{fig1}, but for a linearly polarized standing kink mode. (An animation of this figure is available in the online version)}
        \label{fig2}
\end{figure*}

The simulations evolve for $\approx 70\ \mathrm{min}$, or $\approx 4.5$ oscillation periods in the case of standing waves. We run simulations of both standing and propagating waves, both circularly and linearly polarized, and also low (linear) and high (nonlinear) amplitude versions. Therefore, comparisons of damping times, oscillations periods, transverse-wave induced Kelvin-Helmholtz (TWIKH) growth rates, etc., can be made between the circularly and linearly polarized setups. First we present the results of low amplitude standing wave simulations. We set the amplitude to $A = 3.17\ \mathrm{\mu Pa}$, which results in a maximal displacement of the loop of roughly $0.025a$. Kink oscillations of coronal loops can be considered to evolve in the linear regime when the ratio of maximal displacement to loop radius is much less than unity \citep{2014SoPh..289.1999R}, which is satisfied in this case. The evolution of the circularly polarized standing kink wave is presented\footnote{in the online animated version} in Fig.~\ref{fig1}.
In order to emphasize the dynamics in the inhomogeneous layer, where resonant absorption is taking place, for the simulation in Figure~\ref{fig1} a thick inhomogeneous layer ($l=a$) combined with a diffusive, \texttt{hll} solver is employed. The flow of energy from the core region of the tube to the inhomogeneous layer, also called resonant absorption, is clearly visible in Fig.~\ref{fig1}. For comparison, a simulation of the well-known case of linear polarization with the same parameters is shown in Fig.~\ref{fig2}.
In these figures, resonant absorption can be identified as the kinetic energy growth in the inhomogeneous layer, around the resonant position, where the kink speed equals the local Alfv\'en speed. At the same time, the kinetic energy in the core region, which is associated with the kink wave, is decreasing. Resonant absorption is thus a mechanism to transfer energy from the global, large-scale transverse kink oscillation of the flux tube to local azimuthal flows around it.
The differences in evolution between the two cases are obvious: while in the linearly polarized case the usual m=1 torsional Alfv\'en wave pattern is emerging, in the circularly polarized case the resonance leads to spiral patterns. The development of such a spiral appearance is a consequence of phase mixing in the inhomogeneous layer \citep{2015ApJ...803...43S}, as different parts of the kink wave propagate at different speeds within the inhomogeneous layer. Perturbations towards the outer edge of the flux tube overtake the inner edge ones, which in a circularly polarized case results in an increasingly wound up spiral over time. This is in analogy with the linearly polarized case, where over time it leads to a wavy velocity perturbation pattern around the resonant layer \citep[e.g.,][]{1990CoPhC..59...95P,2015ApJ...803...43S}. Although the dynamics of resonant absorption look different, we find that the resonant damping of the kink wave proceeds at the same rate, irrespective of the polarization, which just confirms the expectation from analytics. We have verified this result for three different inhomogeneous layer thicknesses. As in previous studies \citep{2016A&A...595A..81M}, we find that for low amplitudes theoretical damping rates due to resonant absorption \citep{2002ApJ...577..475R,2002ESASP.505..137G} constrain well the measured damping. The measured oscillation period of $\approx 16.2$ min is within 5 per cent of the  analytically predicted one, and independent of the polarization state, as expected. \par
\begin{figure*}[h]
    \centering     
        \includegraphics[width=0.75\textwidth]{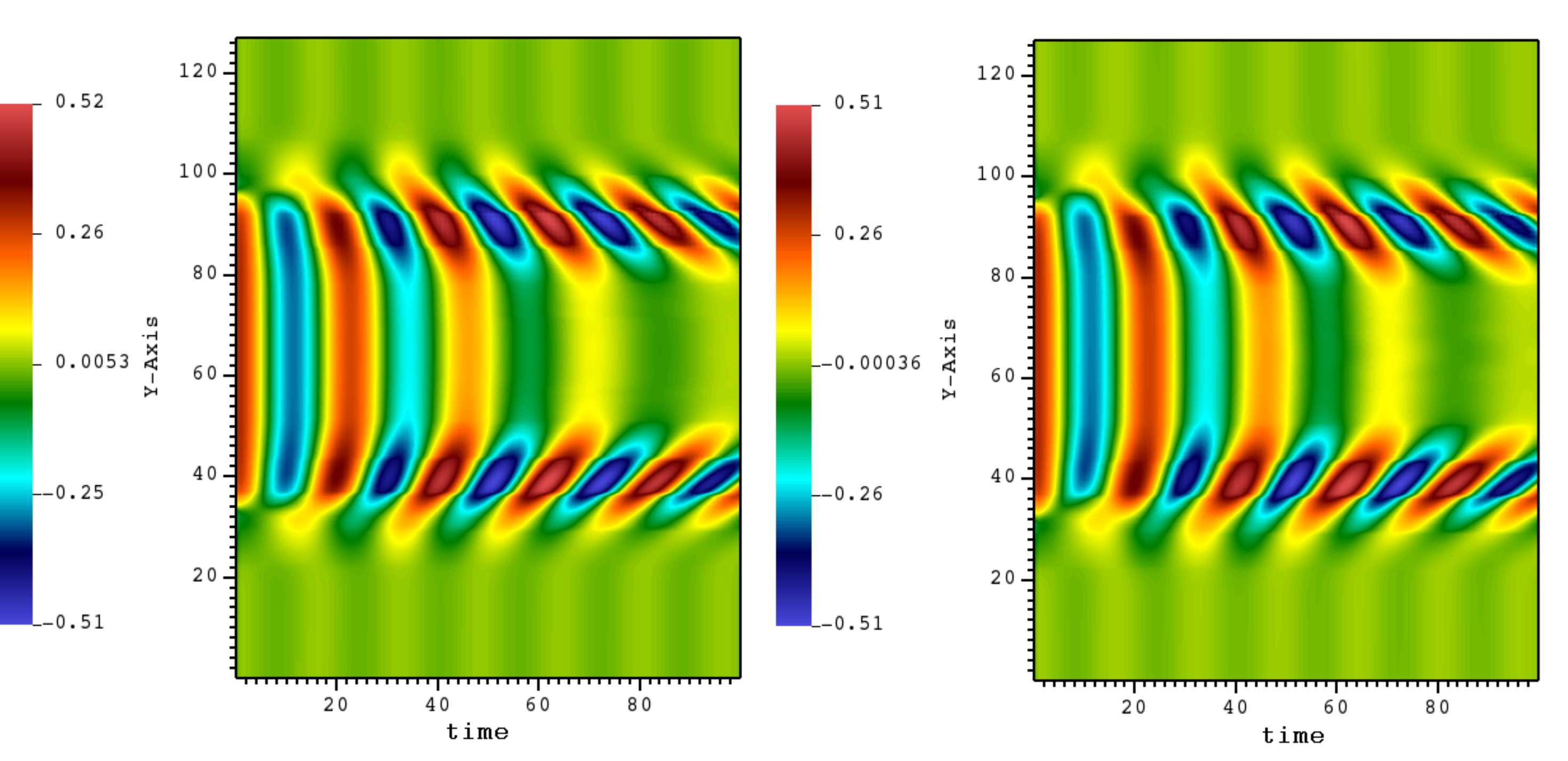} 
        \caption{Time-distance plots of Doppler signal integrated through a slit along the $y$-axis at the mid point of the flux tube, with the LOS the positive x-axis. Left: Linearly polarized kink. Right: Circularly polarized kink. Doppler shift is in code units. Length and time units are $\approx 30$ km and $42$ s, respectively.}
        \label{fig3}
\end{figure*}

In order to determine whether circularly polarized kink waves are distinguishable from linearly polarized ones through Doppler shift measurements, we have forward modeled the simulations, obtaining the emission from the data cubes corresponding to the AIA $171\AA$ channel, shown in Fig.~\ref{fig3}. 
It can be observed that despite resonant absorption appearing differently for a circularly polarized kink wave than for a linearly polarized one, this is not detectable in the Doppler signal. The only tell-tale sign of the circular polarization, albeit small due to the small displacement amplitude, is the alternating vertical displacement of the Doppler signals, which is absent in the linearly polarized case. Note however that while the Doppler signal is independent of the LOS for the circularly polarized kink, it varies for the linearly polarized one. Some additional discussion on the detectability of circularly polarized kink waves is presented in the next section. \par
\begin{figure*}
    \centering
    \includegraphics[width=0.75\textwidth]{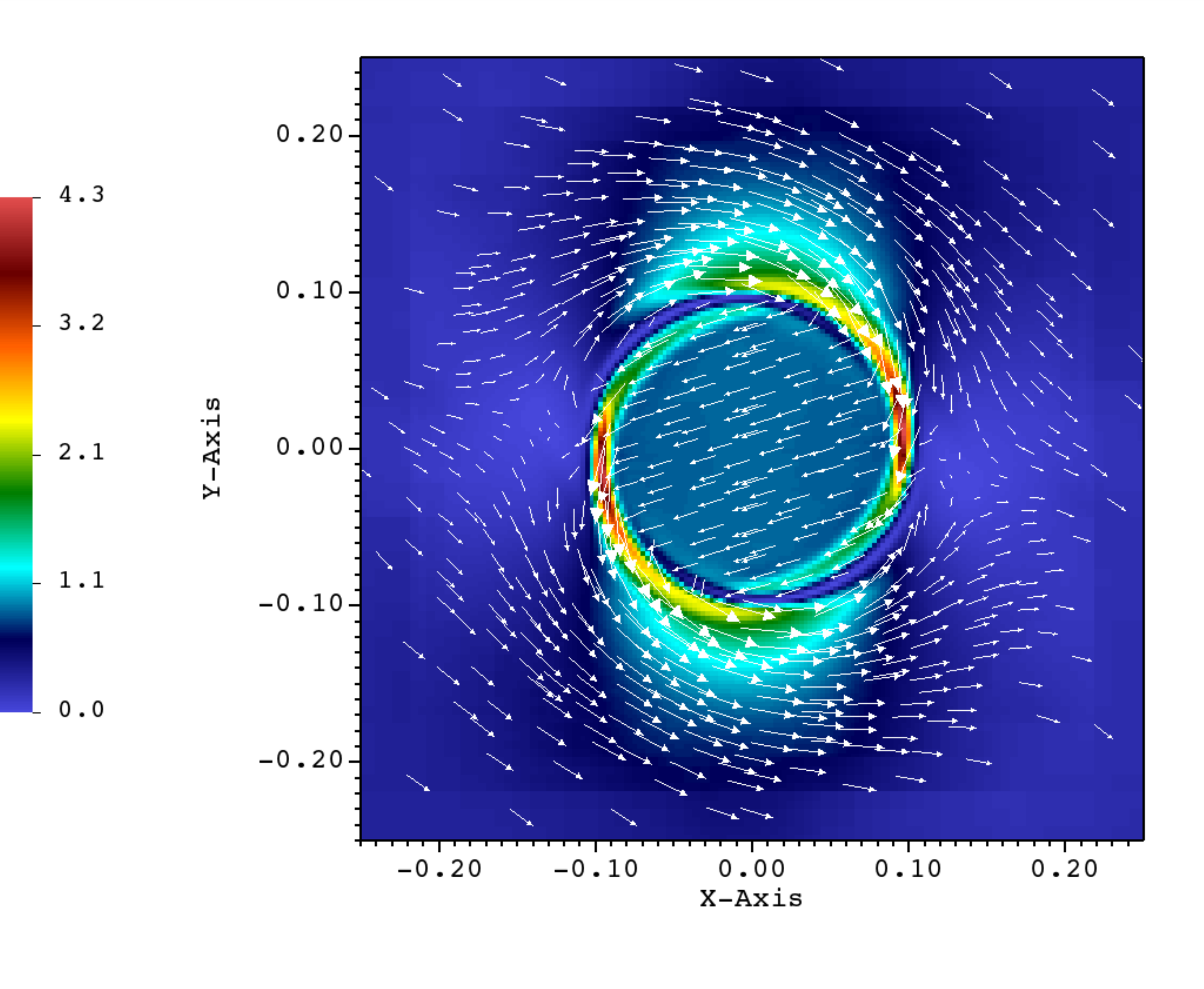} 
    \caption{Plot of kinetic energy and velocity vectors in the cross-sectional slice at $z = 50$ Mm, for the propagating circularly polarized kink wave. Axis units are 10 Mm. Kinetic energy is in code units. The size of the velocity vectors is proportional to their magnitude.}
    \label{figprop}
\end{figure*}

The case of propagating circularly polarized kink waves in the linear regime naturally shares many similarities to the standing wave, which can be interpreted as the superposition of two counter-propagating waves. In Fig.~\ref{figprop} the cross-sectional kink wave dynamics for a flux tube with $l=0.25a$ is shown. 
It is worth noting that in the propagating wave setup, the velocity perturbation that drives the kink wave at the bottom boundary of the numerical domain does not depend on position. That is, the whole boundary is driven. Therefore, while the driver excites propagating kink waves in the flux tube, in the homogeneous medium surrounding the flux tube there are also propagating circularly polarized Alfv\'en waves excited. These waves propagate then in superposition through the numerical domain. The kink wave solution falls off exponentially outside the flux tube, leaving the Alfv\'en waves appearing dominant sufficiently far away from it. These can be seen as the velocity perturbations in the homogeneous region outside of the flux tube in Fig~\ref{figprop} (compare to Fig.~\ref{fig1} for an estimation of the exponential scale length of the perturbation outside the flux tube). 
In a similar fashion to the standing wave, we have compared the Doppler signals of propagating, linearly and circularly polarized kink waves. The result is very similar to the one in Fig.~\ref{fig3}, albeit the time axis gets replaced by the $z$-axis, meaning that the difficulty in detecting kink wave polarization from Doppler signals is still valid for propagating waves.
\par
\begin{figure*}[h]
    \centering     
        \includegraphics[width=0.75\textwidth]{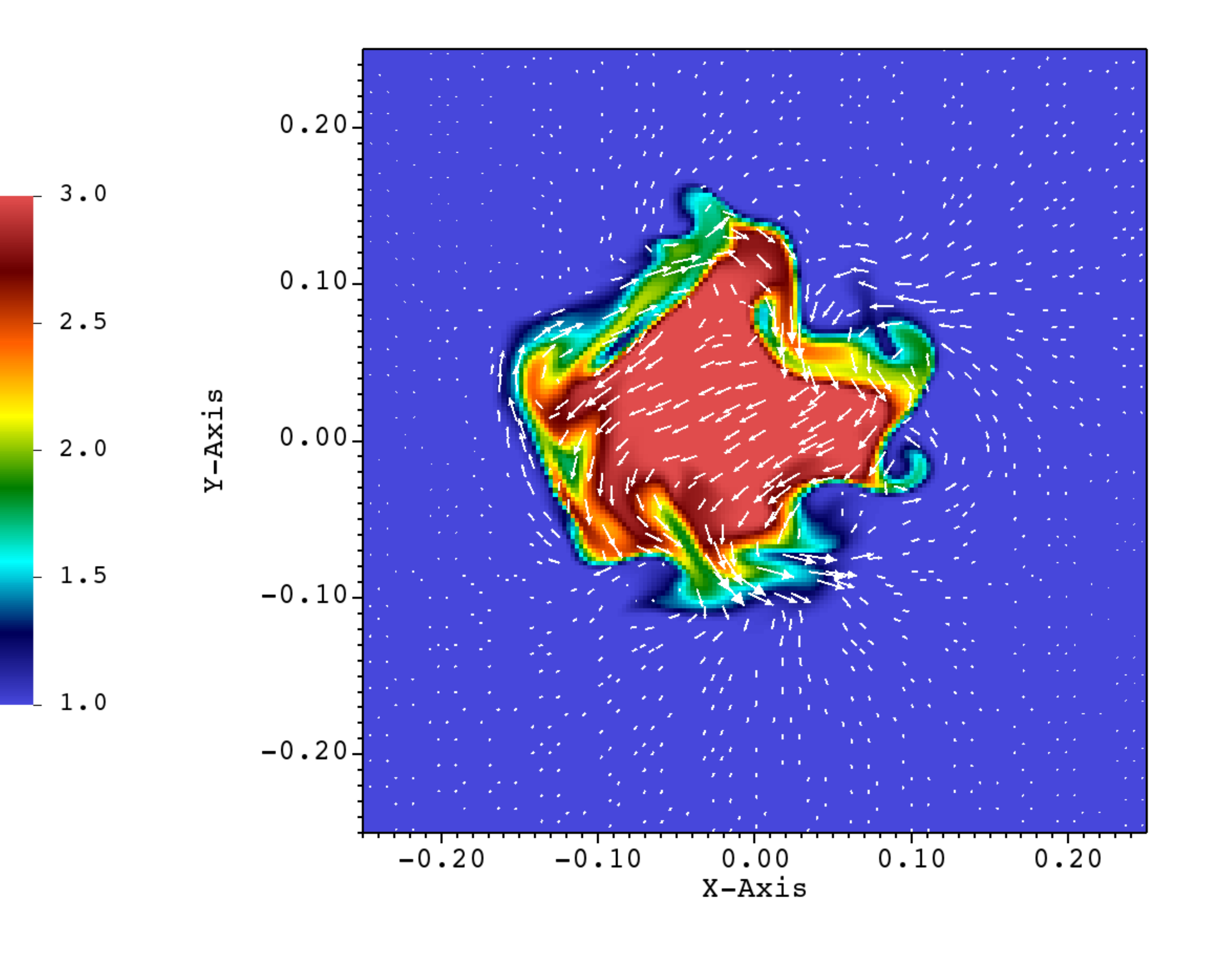} 
        \caption{Plot of density and velocity vectors in the cross-sectional slice at the midpoint of the flux tube, for the nonlinear setup, after approximately two fundamental kink mode periods of evolution. Axis units are 10 Mm. Density is in code units. The size of the velocity vectors is proportional to their magnitude. (An animation of this figure is available in the online version)}
        \label{fig4}
\end{figure*}

\begin{figure*}[h]
    \centering     
        \includegraphics[width=0.75\textwidth]{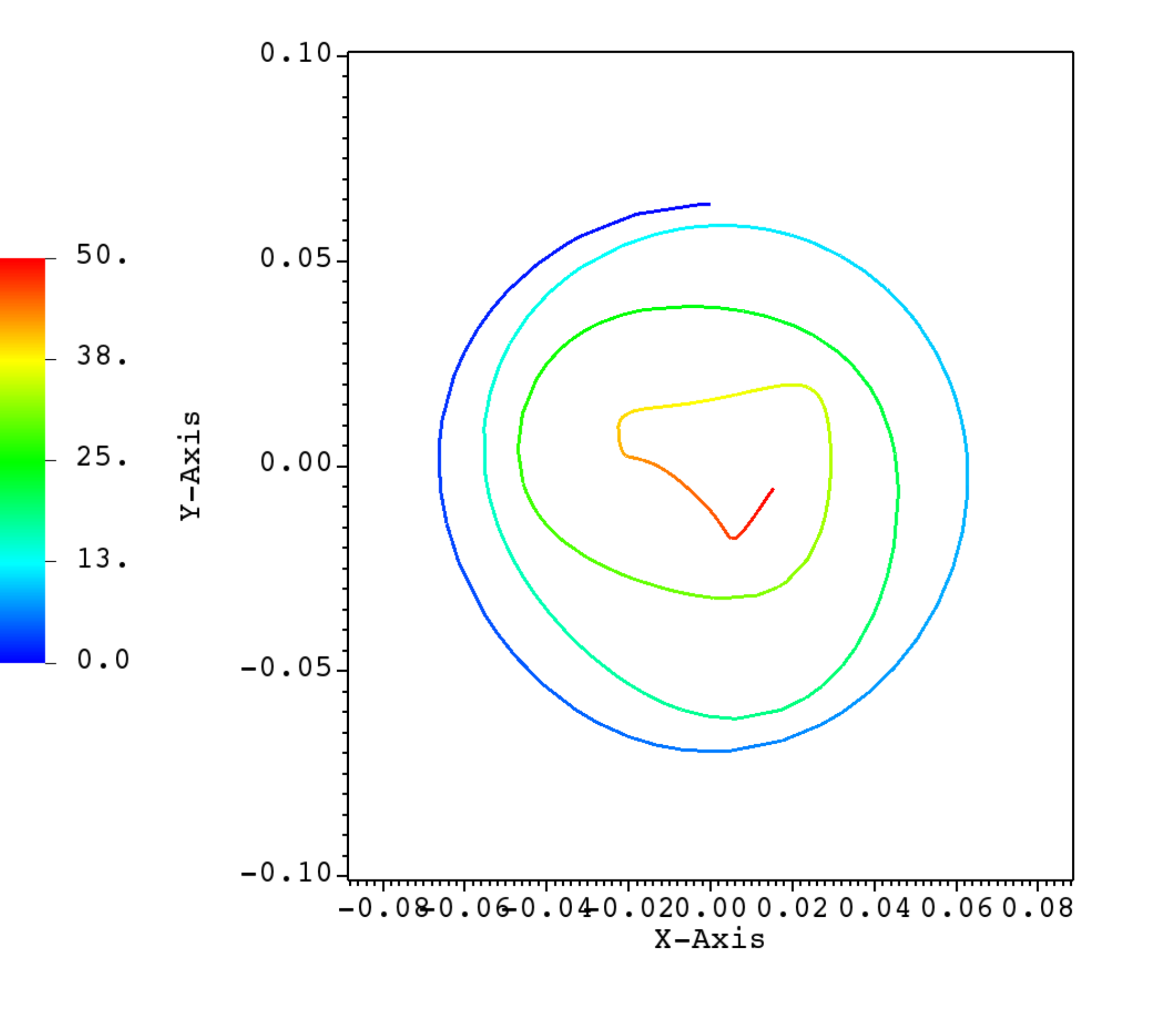} 
        \caption{Path line plot obtained by integrating the velocity map over the duration of the simulation in the tube cross-section, showing the displacement of the flux tube midpoint over time. The color corresponds to the snapshot number, thus showing advance in time, with "0" the initial condition and "50" denoting the final snapshot. Axis units are 10 Mm.}
        \label{fig5}
\end{figure*}

\begin{figure*}[h]
    \centering     
        \includegraphics[width=1.0\textwidth]{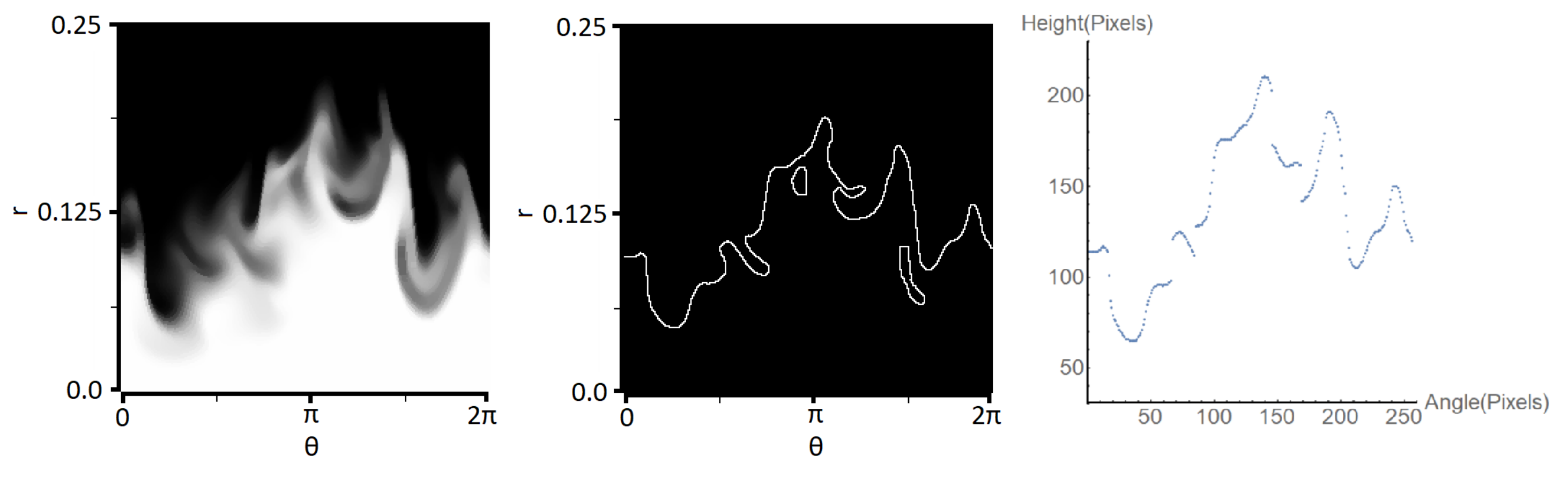} 
        \caption{Left: Polar transform of the density plot shown in Fig.~\ref{fig4}, smoothed using the routine \texttt{MedianFilter} in \texttt{Mathematica}. Centre: The flux tube boundary identified from the polar transformed plot, using the routine \texttt{MorphologicalPerimeter} in \texttt{Mathematica}. Right: The data points extracted from the flux tube boundary plot. Radius units are 10 Mm. }
        \label{fig4a}
\end{figure*}

\begin{figure*}[t]
    \centering     
     \begin{tabular}{@{}cc@{}}
        \includegraphics[width=0.5\textwidth]{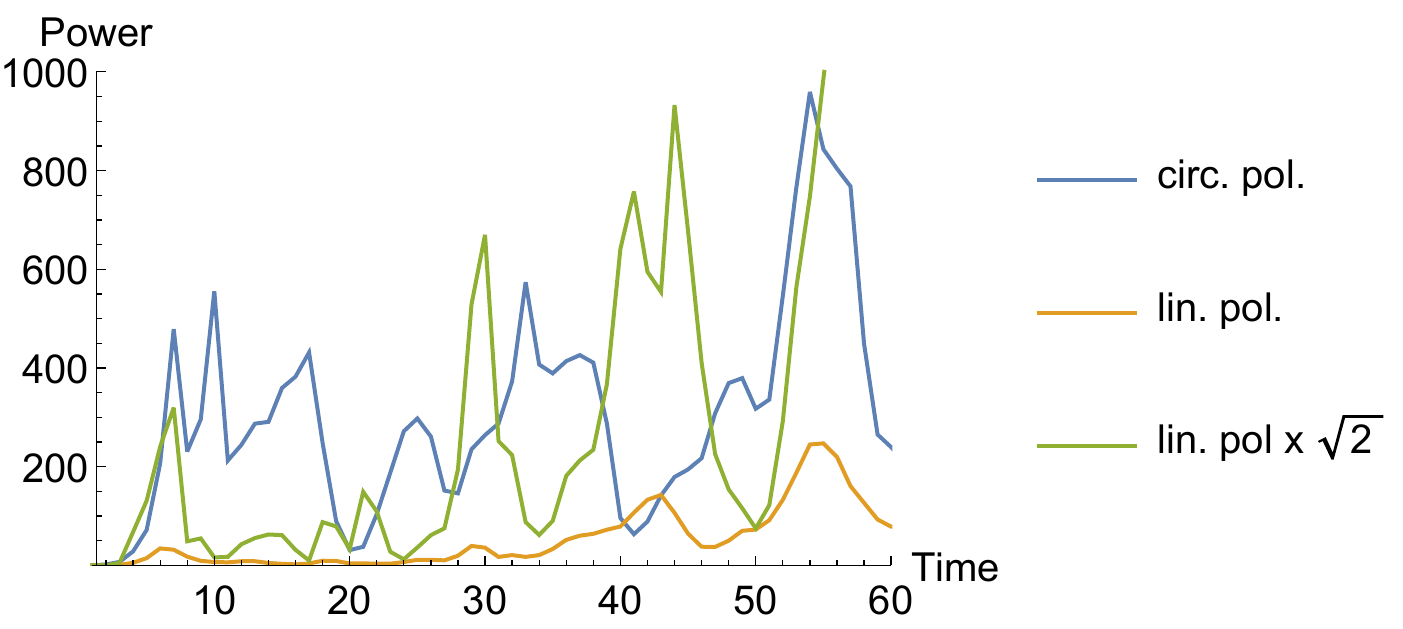}  
        \includegraphics[width=0.5\textwidth]{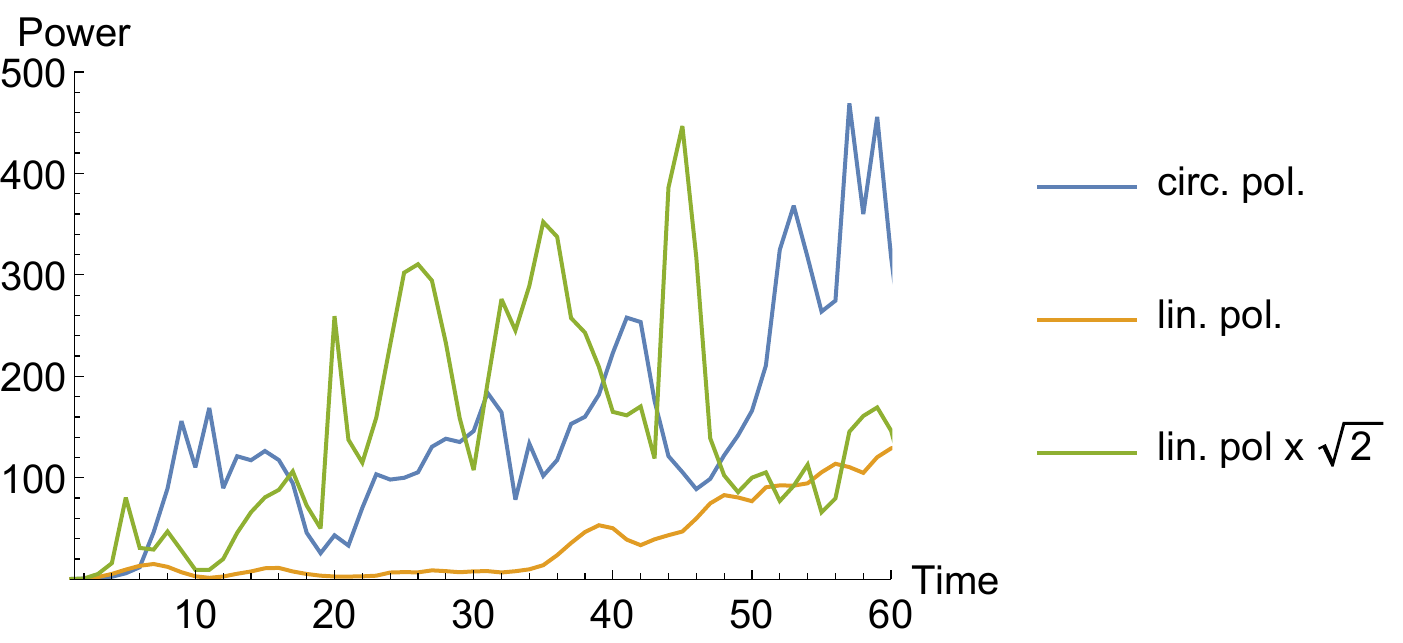} \\
       \end{tabular}    
        \caption{Plots showing the summed contribution of higher angular harmonics, summing over the Fourier components in a lower range, i.e. $\sum_{m=3}^{5} |G(m)|^2$ (left) and higher range $\sum_{m=6}^{15} |G(m)|^2$ (right), as a function of time, in units of $\approx 85.9$ s (duration between snapshots).}
        \label{fig6}
\end{figure*}

 
In the following, the nonlinear evolution of the circularly polarized kink wave will be investigated. The amplitude is set to  $A = 60.34\ \mathrm{\mu Pa}$, which leads to a displacement of $\approx 0.6 a$, in the nonlinear regime. The inhomogeneous layer is set to be defined only by the diffusivity of the numerical solver, thus setting initially $l = 0$. This ensures that nonlinear processes at the edge of the flux tube develop rapidly \citep[see, e.g.][]{2018ApJ...853...35T}. In Fig.~\ref{fig4}, the nonlinear dynamics are shown in the kink anti-node cross-section. 
As it is well known from simulations on the nonlinear evolution of the linearly polarized kink \citep[e.g.,][and many others]{2008ApJ...687L.115T,2014ApJ...787L..22A,2015A&A...582A.117M}, at the edge of the flux tube TWIKH-rolls develop. These roll-ups appear less symmetric than in a linearly polarized situation, with the asymmetry inducing deviations from circular polarization.  This is illustrated in Fig.~\ref{fig5}, where the trajectory of the flux tube midpoint is traced for the whole duration of the nonlinear simulation.
Note that initially the trajectory is close to circular, albeit damping leading to an inward spiral. Once TWIKH rolls develop, the kink oscillation transitions into elliptical polarization, with a more pronounced oscillation along the $x$-axis. Although both numerical \citep[e.g.,][see also above]{2018ApJ...853...35T} and analytical \citep[e.g.][]{2019ApJ...870..108B} works investigated the growth rate of the TWIKH rolls, these studies considered the linearly polarized case in which the shear flow is sinusoidal and it occurs at the same position with respect to the flux tube. In the case of a circularly polarized kink wave, at least in the idealized linear regime with discontinuous boundary, the perturbation amplitude is constant, and it undergoes rotation with respect to the flux tube. Thus, from a Lagrangian perspective (i.e. following the motion of the flux tube during the oscillation), at a particular position along the flux tube boundary, the shear still undergoes a sinusoidal evolution. However, the difference compared to the linearly polarized case is that the shear is present all around the boundary within an oscillation period. This has implications on the growth rate of higher angular harmonics, specific to the development of TWIKH rolls. Moreover, the integrated energy over an oscillation period is 2 times higher for the same amplitude for a circularly polarized kink wave compared to a linearly polarized oscillation. Therefore, in order to test the effect of circular polarization on the growth of TWIKH rolls, the growth rate is compared to simulations with linear polarization of both the same amplitude and $\sqrt{2}$ times the amplitude. In Fig.~\ref{fig6}, the power in high angular harmonics is shown, for the linearly and circularly polarized simulations. The method of obtaining the power spectra of the tube perturbation is the following. Starting from the cross-sectional density plot in the kink anti-node, a polar transformation is applied to the image. In the polar image of the cross-section, an undisturbed loop appears as a straight line at the constant radial distance $r=a$, while a kink wave of linear amplitude as a sinusoidal curve fitting exactly one wavelength in the angular direction from 0 to $2 \pi$. Then, using smoothing and pattern identification methods, in-built in \texttt{Mathematica}\footnote{https://www.wolfram.com/mathematica/},  the curve delimiting the tube boundary ($r$ as a function of $\theta$) is obtained, to which we apply the Fourier transform (see Fig~\ref{fig4a}).
Smoothing removes the step-like appearance of the tube boundary due to the limited numerical resolution, and helps in the detection of the boundary. Note that the boundary extraction is complicated by the smooth variation of the density and the generation of TWIKH rolls, which result in local density minima, and are recognized as small closed-curve patches. As the extracted boundary position as a function of the angle cannot be multi-valued, at places where roll-ups appear the point with the largest radius value is chosen. This results in discontinuities appearing at these positions in the extracted curve, altering the relative contribution of highly developed roll-ups to the power spectra. Nevertheless, this technique still allows for the qualitative comparison of the growth rate of small-scales for the two polarizations. For Fig.~\ref{fig6}, we separate the higher amplitude harmonics into a lower ($m \approx 3-5$) and upper range ($m \approx 6-15$), which are summed up in these ranges, respectively. This is justified by the appearance of comparatively larger eddies in the circularly polarized case initially.  The Fourier coefficients are defined as following \citep{2018ApJ...853...35T},
\begin{equation}
    G(m) = \frac{1}{N} \sum_{n=0}^{N-1} g(n) \exp^{-i \frac{2 \pi}{N} n m},
    \label{fourier}
\end{equation}
where N is the discrete total number of data points (n = 0, ..., N-1) extracted along the loop boundary, and g(n) is the extracted boundary curve (r as a function of $\theta$). We define $\theta_n = 2 \pi n/N$ for convenience. Then, the contribution of each mode $m$ to the total signal is given by the inverse Fourier transform,
\begin{equation}
    g(\theta_n) = \sum_{m=0}^{N-1} G(m) \exp^{i m \theta_n}
    \label{invfourier}
\end{equation}
The obtained spectra appear rapidly varying and noisy. Besides the inherently noisy nature of the discrete Fourier transform \citep[e.g.,][]{2005A&A...431..391V}, the extraction process of the boundary curve is thought to account for some of this variability. Therefore, care should be taken when interpreting the spectral evolution presented in Fig.~\ref{fig6}. However, it is clear that in the linearly polarized case with the same amplitude as in the circularly polarized case, higher harmonics develop considerably slower, while for the linearly polarized case with $\sqrt{2}$ times the amplitude the spectral evolution is qualitatively similar to that for the circularly polarized case.
Not shown in Fig~\ref{fig6} but visible\footnote{in the online animated version} in Fig~\ref{fig4}, the coupling to the m=2 fluting mode is also present, as in the linearly polarized case \citep[e.g.,][]{2016A&A...595A..81M}, being especially prominent during the first oscillation period, manifesting as an area-conserving elliptical deformation of the circular cross-section of the tube. A nonlinear propagating wave simulation was also carried out. In this case, the expected deformation and transition to turbulence of the unidirectionally propagating kink wave is observed \citep{2017NatSR...14820M,2019ApJ...873...56M}. The differences between the circularly polarized and the linearly polarized simulation are essentially the same as for the standing wave, in the sense that kink wave energy is a better indicator of nonlinear growth rate than the velocity amplitude of the wave. 

\section{Conclusions} \label{four}

We have studied the circularly polarized kink wave in a cylindrical flux tube through numerical simulations. Both standing and propagating modes were studied, for perturbation amplitudes both in the linear and nonlinear regime. The most striking difference between the circularly and linearly polarized cases is the appearance of the wave perturbation, once resonant absorption and phase mixing is starting to act. While in the linearly polarized case the resonant flow at the edge of the tube is in the direction of the transverse kink perturbation, in the circularly polarized case it is spiral-like and can be perpendicular to the kink perturbation. Although resonant absorption shows morphological differences for linear and circular polarization, no impact on the damping time or oscillation period was found, in agreement with analytical results. In the case of propagating waves, the same conclusions can be drawn on the differences between the circularly and linearly polarized kink waves, with variation in time replaced by variation over propagation distance for continuously driven waves. \par In the nonlinear regime, for displacements comparable to the tube radius, a tube undergoing circularly polarized kink oscillations deforms and develops small scales or TWIKH rolls faster than a linearly polarized kink wave with the same amplitude. We find that the growth rate for nonlinear small scale generation is governed by the energy of the oscillation rather than the perturbation amplitude. Therefore, a growth rate comparable to the one in the circularly polarized kink was achieved with a linearly polarized kink wave with $\sqrt{2}$ times the former's amplitude. This is also valid for nonlinear propagating kink waves. \par
An important point is the observability of the polarization state of kink waves in the corona. We find that despite the previously described differences between the appearance of resonant flows, there is no difference in the Doppler signal. The only tell-tale signature of circular polarization in the studied case is the displacement of alternating Doppler signals. However, note that in the linear case there would be no displacement observed only if the LOS is parallel to the polarization direction. The Doppler signal is independent of LOS direction in the circularly polarized case. In the linearly polarized case, while the Doppler signal qualitatively remains the same for different LOS directions, the amplitude varies roughly with the cosine of the angle between LOS and polarization direction \citep[see, e.g.][]{2016ApJS..223...23Y}. In the near future, simultaneous observations of coronal loop oscillations by the Extreme Ultraviolet Imager on board the Solar Orbiter mission, and from SDO/AIA, could provide the opportunity to perform stereoscopic analysis with unprecedented resolution. 
With suitable vantage points (i.e. spacecraft separation angles $\leq 90^{\circ}$) the complete 3D structure of the loop may be deduced without the limiting assumption of semi-circularity.
Then with sufficiently long-lived data, the polarization of the transverse oscillation with respect to the (modeled) plane of the loop may be calculated through comparison of the projected/apparent displacement in the two fields of view. 
The ubiquity of decayless kink oscillations in the solar corona implies that there will be at least some targets for stereoscopic analysis. 
Curiously, all known detections of decayless oscillations have been polarized in the horizontal direction, as with the large amplitude regime \citep{2015A&A...583A.136A}. 
Whether this is an observational bias caused by the studies' single field of view has important implications for the driver, and (apparent lack of) damping mechanisms of decayless oscillations. 
We note that even if all kink oscillations are initially horizontally polarized, the expansion or contraction of the hosting coronal loop \citep[a common occurrence, e.g.][]{2015A&A...581A...8R} may still provide the opportunity to observe transverse oscillations which are elliptically polarized. 
We suggest that observations looking for slowly expanding/contracting coronal loops exhibiting transverse kink oscillations, observed by both Solar Orbiter and SDO, may be used to substantiate the results presented here.  \par 
The caveats of this study are the simplifications used in modeling coronal loops. We neglect the curvature of the loops, opting for a straight flux tube. Nevertheless, \citet{2004A&A...424.1065V} have demonstrated that loop curvature has a minimal effect on the kink eigenmodes and the differences between horizontal and vertical polarization are negligible, although \citet{2009A&A...506..885R} points out that this is only valid when the loop expansion (varying cross-section along the loop) is neglected, as in the present study. We also note that \citet{2020ApJ...904..116G} have shown that any ellipticity of the loop cross-section affects both the eigenfrequencies and the damping rates by resonant absorption for different polarizations.
Gravity is also neglected, leading to a homogeneous density and Alfv\'en speed profile along the loop. The density scale height in the corona is around $50\ \mathrm{Mm}$ for a $1\ \mathrm{MK}$ plasma, leading to large variations in Alfv\'en speed within long loops. While the ratios of periods of different standing kink wave harmonics are affected by stratification \citep[e.g.][]{2005ApJ...624L..57A}, we do not expect a polarization-dependent effect. We have only studied two very specific polarization states, linear and circular. It seems improbable that various coronal mechanisms excite exactly these specific polarization states of the kink oscillation, with most oscillations probably in an elliptical polarization state. However, as elliptical polarization can be positioned between the two extremes of linear and circular polarization, our conclusions on the differences or similarities between these two states should still hold for the elliptical polarization state. \par 
In the following, we discuss on the potential applicability of this study. The polarization state of observed kink oscillations of coronal loops is often not determined, with a few exceptions noted in the Introduction. However, the polarization state could offer us important seismological information on the driving mechanism of kink oscillations, especially in the context of small amplitude decayless oscillations \citep[e.g.][]{2013A&A...552A..57N,2015A&A...583A.136A}, which are thought to be continuously reinforced standing kink oscillations. In this sense, a statistical analysis on the polarization state of these oscillations could constrain some properties of the loop footpoint driver, such as variability, amount of vorticity, and time correlation. In particular, some explanations of decayless kink oscillation arising as a self-oscillatory process \citep{2016A&A...591L...5N} entail that the flow inducing the oscillation is quasi-stationary, which would lead to linearly polarized oscillations.

\bibliography{Biblio}

\begin{thebibliography}{47}
\expandafter\ifx\csname natexlab\endcsname\relax\def\natexlab#1{#1}\fi

\bibitem[{{Andries} {et~al.}(2005){Andries}, {Arregui}, \&
  {Goossens}}]{2005ApJ...624L..57A}
{Andries}, J., {Arregui}, I., \& {Goossens}, M. 2005, \apjl, 624, L57

\bibitem[{{Anfinogentov} {et~al.}(2015){Anfinogentov}, {Nakariakov}, \&
  {Nistic{\`o}}}]{2015A&A...583A.136A}
{Anfinogentov}, S.~A., {Nakariakov}, V.~M., \& {Nistic{\`o}}, G. 2015, \aap,
  583, A136

\bibitem[{{Antolin} {et~al.}(2014){Antolin}, {Yokoyama}, \& {Van
  Doorsselaere}}]{2014ApJ...787L..22A}
{Antolin}, P., {Yokoyama}, T., \& {Van Doorsselaere}, T. 2014, \apjl, 787, L22

\bibitem[{{Arregui}(2015)}]{2015RSPTA.37340261A}
{Arregui}, I. 2015, Philosophical Transactions of the Royal Society of London
  Series A, 373, 20140261

\bibitem[{{Arregui}(2018)}]{2018AdSpR..61..655A}
{Arregui}, I. 2018, Advances in Space Research, 61, 655

\bibitem[{{Arregui}(2021)}]{2021ApJ...915L..25A}
{Arregui}, I. 2021, \apjl, 915, L25

\bibitem[{{Aschwanden}(2009)}]{2009SSRv..149...31A}
{Aschwanden}, M.~J. 2009, \ssr, 149, 31

\bibitem[{{Barbulescu} {et~al.}(2019){Barbulescu}, {Ruderman}, {Van
  Doorsselaere}, \& {Erd{\'e}lyi}}]{2019ApJ...870..108B}
{Barbulescu}, M., {Ruderman}, M.~S., {Van Doorsselaere}, T., \& {Erd{\'e}lyi},
  R. 2019, \apj, 870, 108

\bibitem[{{De Moortel} \& {Nakariakov}(2012)}]{2012RSPTA.370.3193D}
{De Moortel}, I. \& {Nakariakov}, V.~M. 2012, Royal Society of London
  Philosophical Transactions Series A, 370, 3193

\bibitem[{{Edwin} \& {Roberts}(1983)}]{1983SoPh...88..179E}
{Edwin}, P.~M. \& {Roberts}, B. 1983, \solphys, 88, 179

\bibitem[{{Goddard} \& {Nakariakov}(2016)}]{2016A&A...590L...5G}
{Goddard}, C.~R. \& {Nakariakov}, V.~M. 2016, \aap, 590, L5

\bibitem[{{Goossens} {et~al.}(2002){Goossens}, {de Groof}, \&
  {Andries}}]{2002ESASP.505..137G}
{Goossens}, M., {de Groof}, A., \& {Andries}, J. 2002, in ESA Special
  Publication, Vol. 505, SOLMAG 2002. Proceedings of the Magnetic Coupling of
  the Solar Atmosphere Euroconference, ed. H.~{Sawaya-Lacoste}, 137--144

\bibitem[{{Guo} {et~al.}(2020){Guo}, {Li}, \& {Van
  Doorsselaere}}]{2020ApJ...904..116G}
{Guo}, M., {Li}, B., \& {Van Doorsselaere}, T. 2020, \apj, 904, 116

\bibitem[{{Jess} {et~al.}(2017){Jess}, {Van Doorsselaere}, {Verth}, {Fedun},
  {Krishna Prasad}, {Erd{\'e}lyi}, {Keys}, {Grant}, {Uitenbroek}, \&
  {Christian}}]{2017ApJ...842...59J}
{Jess}, D.~B., {Van Doorsselaere}, T., {Verth}, G., {et~al.} 2017, \apj, 842,
  59

\bibitem[{{Magyar} \& {Van Doorsselaere}(2016)}]{2016A&A...595A..81M}
{Magyar}, N. \& {Van Doorsselaere}, T. 2016, \aap, 595, A81

\bibitem[{{Magyar} {et~al.}(2019){Magyar}, {Van Doorsselaere}, \&
  {Goossens}}]{2019ApJ...873...56M}
{Magyar}, N., {Van Doorsselaere}, T., \& {Goossens}, M. 2019, \apj, 873, 56

\bibitem[{{Magyar} {et~al.}(2017){Magyar}, {Van Doorsselaere}, \&
  {Gossens}}]{2017NatSR...14820M}
{Magyar}, N., {Van Doorsselaere}, T., \& {Gossens}, M. 2017, Scientific
  Reports, 7, 14820

\bibitem[{{Magyar} {et~al.}(2015){Magyar}, {Van Doorsselaere}, \&
  {Marcu}}]{2015A&A...582A.117M}
{Magyar}, N., {Van Doorsselaere}, T., \& {Marcu}, A. 2015, \aap, 582, A117

\bibitem[{{Nakariakov} {et~al.}(2021){Nakariakov}, {Anfinogentov}, {Antolin},
  {Jain}, {Kolotkov}, {Kupriyanova}, {Li}, {Magyar}, {Nistic{\`o}}, {Pascoe},
  {Srivastava}, {Terradas}, {Vasheghani Farahani}, {Verth}, {Yuan}, \&
  {Zimovets}}]{2021SSRv..217...73N}
{Nakariakov}, V.~M., {Anfinogentov}, S.~A., {Antolin}, P., {et~al.} 2021, \ssr,
  217, 73

\bibitem[{{Nakariakov} {et~al.}(2016){Nakariakov}, {Anfinogentov},
  {Nistic{\`o}}, \& {Lee}}]{2016A&A...591L...5N}
{Nakariakov}, V.~M., {Anfinogentov}, S.~A., {Nistic{\`o}}, G., \& {Lee}, D.~H.
  2016, \aap, 591, L5

\bibitem[{{Nakariakov} \& {Kolotkov}(2020)}]{2020ARA&A..58..441N}
{Nakariakov}, V.~M. \& {Kolotkov}, D.~Y. 2020, \araa, 58, 441

\bibitem[{{Nechaeva} {et~al.}(2019){Nechaeva}, {Zimovets}, {Nakariakov}, \&
  {Goddard}}]{2019ApJS..241...31N}
{Nechaeva}, A., {Zimovets}, I.~V., {Nakariakov}, V.~M., \& {Goddard}, C.~R.
  2019, \apjs, 241, 31

\bibitem[{{Nistic{\`o}} {et~al.}(2013){Nistic{\`o}}, {Nakariakov}, \&
  {Verwichte}}]{2013A&A...552A..57N}
{Nistic{\`o}}, G., {Nakariakov}, V.~M., \& {Verwichte}, E. 2013, \aap, 552, A57

\bibitem[{{Poedts} {et~al.}(1990){Poedts}, {Goossens}, \&
  {Kerner}}]{1990CoPhC..59...95P}
{Poedts}, S., {Goossens}, M., \& {Kerner}, W. 1990, Computer Physics
  Communications, 59, 95

\bibitem[{{Porth} {et~al.}(2014){Porth}, {Xia}, {Hendrix}, {Moschou}, \&
  {Keppens}}]{2014ApJS..214....4P}
{Porth}, O., {Xia}, C., {Hendrix}, T., {Moschou}, S.~P., \& {Keppens}, R. 2014,
  \apjs, 214, 4

\bibitem[{{Raymond} {et~al.}(2014){Raymond}, {McCauley}, {Cranmer}, \&
  {Downs}}]{2014ApJ...788..152R}
{Raymond}, J.~C., {McCauley}, P.~I., {Cranmer}, S.~R., \& {Downs}, C. 2014,
  \apj, 788, 152

\bibitem[{{Ruderman}(2009)}]{2009A&A...506..885R}
{Ruderman}, M.~S. 2009, \aap, 506, 885

\bibitem[{{Ruderman} \& {Erd{\'e}lyi}(2009)}]{2009SSRv..149..199R}
{Ruderman}, M.~S. \& {Erd{\'e}lyi}, R. 2009, \ssr, 149, 199

\bibitem[{{Ruderman} \& {Goossens}(2014)}]{2014SoPh..289.1999R}
{Ruderman}, M.~S. \& {Goossens}, M. 2014, \solphys, 289, 1999

\bibitem[{{Ruderman} \& {Roberts}(2002)}]{2002ApJ...577..475R}
{Ruderman}, M.~S. \& {Roberts}, B. 2002, \apj, 577, 475

\bibitem[{{Ruderman} \& {Terradas}(2015)}]{2015A&A...580A..57R}
{Ruderman}, M.~S. \& {Terradas}, J. 2015, \aap, 580, A57

\bibitem[{{Russell} {et~al.}(2015){Russell}, {Sim{\~o}es}, \&
  {Fletcher}}]{2015A&A...581A...8R}
{Russell}, A.~J.~B., {Sim{\~o}es}, P.~J.~A., \& {Fletcher}, L. 2015, \aap, 581,
  A8

\bibitem[{{Soler} \& {Terradas}(2015)}]{2015ApJ...803...43S}
{Soler}, R. \& {Terradas}, J. 2015, \apj, 803, 43

\bibitem[{{Stangalini} {et~al.}(2017){Stangalini}, {Giannattasio},
  {Erd{\'e}lyi}, {Jafarzadeh}, {Consolini}, {Criscuoli}, {Ermolli},
  {Guglielmino}, \& {Zuccarello}}]{2017ApJ...840...19S}
{Stangalini}, M., {Giannattasio}, F., {Erd{\'e}lyi}, R., {et~al.} 2017, \apj,
  840, 19

\bibitem[{{Terradas} {et~al.}(2008){Terradas}, {Andries}, {Goossens},
  {Arregui}, {Oliver}, \& {Ballester}}]{2008ApJ...687L.115T}
{Terradas}, J., {Andries}, J., {Goossens}, M., {et~al.} 2008, \apjl, 687, L115

\bibitem[{{Terradas} {et~al.}(2018){Terradas}, {Magyar}, \& {Van
  Doorsselaere}}]{2018ApJ...853...35T}
{Terradas}, J., {Magyar}, N., \& {Van Doorsselaere}, T. 2018, \apj, 853, 35

\bibitem[{{Teunissen} \& {Keppens}(2019)}]{2019CoPhC.24506866T}
{Teunissen}, J. \& {Keppens}, R. 2019, Computer Physics Communications, 245,
  106866

\bibitem[{{Van Doorsselaere} {et~al.}(2004){Van Doorsselaere}, {Debosscher},
  {Andries}, \& {Poedts}}]{2004A&A...424.1065V}
{Van Doorsselaere}, T., {Debosscher}, A., {Andries}, J., \& {Poedts}, S. 2004,
  \aap, 424, 1065

\bibitem[{{Van Doorsselaere} {et~al.}(2021){Van Doorsselaere}, {Goossens},
  {Magyar}, {Ruderman}, \& {Ismayilli}}]{2021ApJ...910...58V}
{Van Doorsselaere}, T., {Goossens}, M., {Magyar}, N., {Ruderman}, M.~S., \&
  {Ismayilli}, R. 2021, \apj, 910, 58

\bibitem[{{Van Doorsselaere} {et~al.}(2020{\natexlab{a}}){Van Doorsselaere},
  {Li}, {Goossens}, {Hnat}, \& {Magyar}}]{2020ApJ...899..100V}
{Van Doorsselaere}, T., {Li}, B., {Goossens}, M., {Hnat}, B., \& {Magyar}, N.
  2020{\natexlab{a}}, \apj, 899, 100

\bibitem[{{Van Doorsselaere} {et~al.}(2020{\natexlab{b}}){Van Doorsselaere},
  {Srivastava}, {Antolin}, {Magyar}, {Vasheghani Farahani}, {Tian}, {Kolotkov},
  {Ofman}, {Guo}, {Arregui}, {De Moortel}, \& {Pascoe}}]{2020SSRv..216..140V}
{Van Doorsselaere}, T., {Srivastava}, A.~K., {Antolin}, P., {et~al.}
  2020{\natexlab{b}}, \ssr, 216, 140

\bibitem[{{Vaughan}(2005)}]{2005A&A...431..391V}
{Vaughan}, S. 2005, \aap, 431, 391

\bibitem[{{Wang} {et~al.}(2008){Wang}, {Solanki}, \&
  {Selwa}}]{2008A&A...489.1307W}
{Wang}, T.~J., {Solanki}, S.~K., \& {Selwa}, M. 2008, \aap, 489, 1307

\bibitem[{{Williams} {et~al.}(2020){Williams}, {Walsh}, {Winebarger}, {Brooks},
  {Cirtain}, {De Pontieu}, {Golub}, {Kobayashi}, {McKenzie}, {Morton}, {Peter},
  {Rachmeler}, {Savage}, {Testa}, {Tiwari}, {Warren}, \&
  {Watkinson}}]{2020ApJ...892..134W}
{Williams}, T., {Walsh}, R.~W., {Winebarger}, A.~R., {et~al.} 2020, \apj, 892,
  134

\bibitem[{{Xia} {et~al.}(2018){Xia}, {Teunissen}, {El Mellah}, {Chan{\'e}}, \&
  {Keppens}}]{2018ApJS..234...30X}
{Xia}, C., {Teunissen}, J., {El Mellah}, I., {Chan{\'e}}, E., \& {Keppens}, R.
  2018, \apjs, 234, 30

\bibitem[{{Yuan} \& {Van Doorsselaere}(2016)}]{2016ApJS..223...23Y}
{Yuan}, D. \& {Van Doorsselaere}, T. 2016, \apjs, 223, 23

\bibitem[{{Zaitsev} \& {Stepanov}(1975)}]{1975...37..3}
{Zaitsev}, V.~V. \& {Stepanov}, A.~V. 1975, ssled. Geomagn. Aeron. F iz.
  Solntsa, 37, 3

\end{thebibliography}
\bibliographystyle{aa}

\begin{acknowledgements}
N.M. was supported by a Newton International Fellowship of the Royal Society.
T.D. and T.V.D. were supported by the European Research Council (ERC) under the European Union's Horizon 2020 research and innovation programme (grant agreement No 724326). T.V.D. was also supported by the C1 grant TRACEspace of Internal Funds KU Leuven.
V.M.N. acknowledges the STFC consolidated grant ST/T000252/1 , and the BK21 plus program through the National Research Foundation funded by the Ministry of Education of Korea.
\end{acknowledgements}

\end{document}